\documentclass[amsmath,amssymb,aps, prl,floatfix, twocolumn,mathbbm,superscriptaddress,longbibliography]{revtex4-2}
\usepackage{graphicx,dsfont}
\usepackage[dvipsnames]{xcolor}
\usepackage[colorlinks=true,citecolor=Orange,linkcolor=Orange,urlcolor=Purple]{hyperref}
\usepackage{physics}
\usepackage{bbold}
\usepackage[utf8]{inputenc}
\usepackage[english]{babel}
\usepackage{braket}
\usepackage{mathtools}

\newcommand\givenbase[1][]{\:#1\lvert\:}
\let\given\givenbase

\DeclarePairedDelimiterX\Basics[1](){\let\given\sgiven #1}

\AtBeginDocument{%
	\newwrite\bibnotes
	\def\bibnotesext{Notes.bib}
	\immediate\openout\bibnotes=\jobname\bibnotesext
	\immediate\write\bibnotes{@CONTROL{REVTEX41Control}}
	\immediate\write\bibnotes{@CONTROL{%
			apsrev41Control,author="08",editor="1",pages="1",title="0",year="1"}}
	\if@filesw
	\immediate\write\@auxout{\string\citation{apsrev41Control}}%
	\fi
}%

\begin{document}
\title{Ground-State Cooling of Levitated Magnets in Low-Frequency Traps}
\author{Kirill Streltsov}
\author{Julen S. Pedernales}
\author{Martin B. Plenio}
\affiliation{Institut f\"ur Theoretische Physik und IQST, Albert-Einstein-Allee 11, Universit\"at Ulm, D-89081 Ulm, Germany}
\date{\today}

\begin{abstract}

We present a ground-state cooling scheme for the mechanical degrees of freedom of mesoscopic magnetic particles levitated in low-frequency traps. Our method makes use of a binary sensor and suitably shaped pulses to perform weak, adaptive measurements on the position of the magnet. This allows us to precisely determine the position and momentum of the particle, transforming the initial high-entropy thermal state into a pure coherent state. The energy is then extracted by shifting the trap center. By delegating the task of energy extraction to a coherent displacement operation we overcome the limitations associated with cooling schemes that rely on the dissipation of a two-level system coupled to the oscillator. We numerically benchmark our protocol in realistic experimental conditions, including heating rates and imperfect readout fidelities, showing that it is well suited for magnetogravitational traps operating at cryogenic temperatures. Our results pave the way for ground-state cooling of micron-scale particles.

\end{abstract}

\maketitle

{\it Introduction.---}Cooling the center-of-mass (c.m.) motion of a massive oscillator down to its minimum energy is a convenient way of transferring it from a classical thermal state into a pure quantum state for applications in quantum technologies. The ability to operate massive particles in the quantum regime is predicted to bring exceptional enhancements in sensitivity for metrological applications~\cite{millen2020, moore2021, rademacher2020, weiss2020, cosco2020, libbrecht2004}, and to have relevant implications for our understanding of nature~\cite{bassi2013, arndt2014, geraci2015, rider2016, arvanitaki2013,romero-isart2011, schmole2016, krisnanda2020}. For this enterprise, levitated optomechanics stands out as a promising platform, where nano- and microscale solids suspended in vacuum behave as massive mechanical oscillators. Recently, ground-state (GS) cooling of an optically levitated mesoscopic particle has been demonstrated~\cite{delic2020, magrini2020}, constituting a remarkable milestone for the field. However, optical setups suffer from significant dissipation rates due to recoil and absorption of photons from the trapping fields, which makes progress beyond GS cooling, such as the generation of nonclassical states, daunting. 

In light of these challenges, levitation with passive fields, e.g., magnetogravitational levitation~\cite{prothero1968, goodkind1999, hsu2016, slezak2018, gieseler2020, vinante2020a, leng2021}, which promises extended coherence times, has attracted renewed interest. In these setups the presence of superconductors and the cryogenic operating temperatures~\cite{gieseler2020, vinante2020a, leng2021} make the use of high-intensity laser fields challenging, thereby limiting the applicability of optical cooling schemes. A number of works propose to engineer a low-temperature environment for the oscillator by coupling it to a two-level system (TLS) that is repeatedly initialized in its GS~\cite{rabl2009,lau2016,romero-isart2012,cirio2012}. However, in each initialization cycle, at most one phonon can be extracted, leading to a linear reduction in energy over time and a diminishing cooling rate~\cite{rabl2010} with increasing initial phonon numbers. Therefore, this approach becomes prohibitively slow in passive traps that operate in the low-frequency regime (100 Hz) with initial thermal occupancies of $\bar{n}_{T=300 {\rm\, K}} \approx 10^{10}$ for the c.m. mode at room temperature. Another class of TLS-based cooling schemes relies on the projection of the oscillator onto the GS~\cite{rao2016, montenegro2018, puebla2020}, leading to probabilistic outcomes. These schemes are also limited to low initial phonon numbers as otherwise the success probability becomes extremely small.

\begin{figure}[b]
	\centering
	\includegraphics[width=\columnwidth]{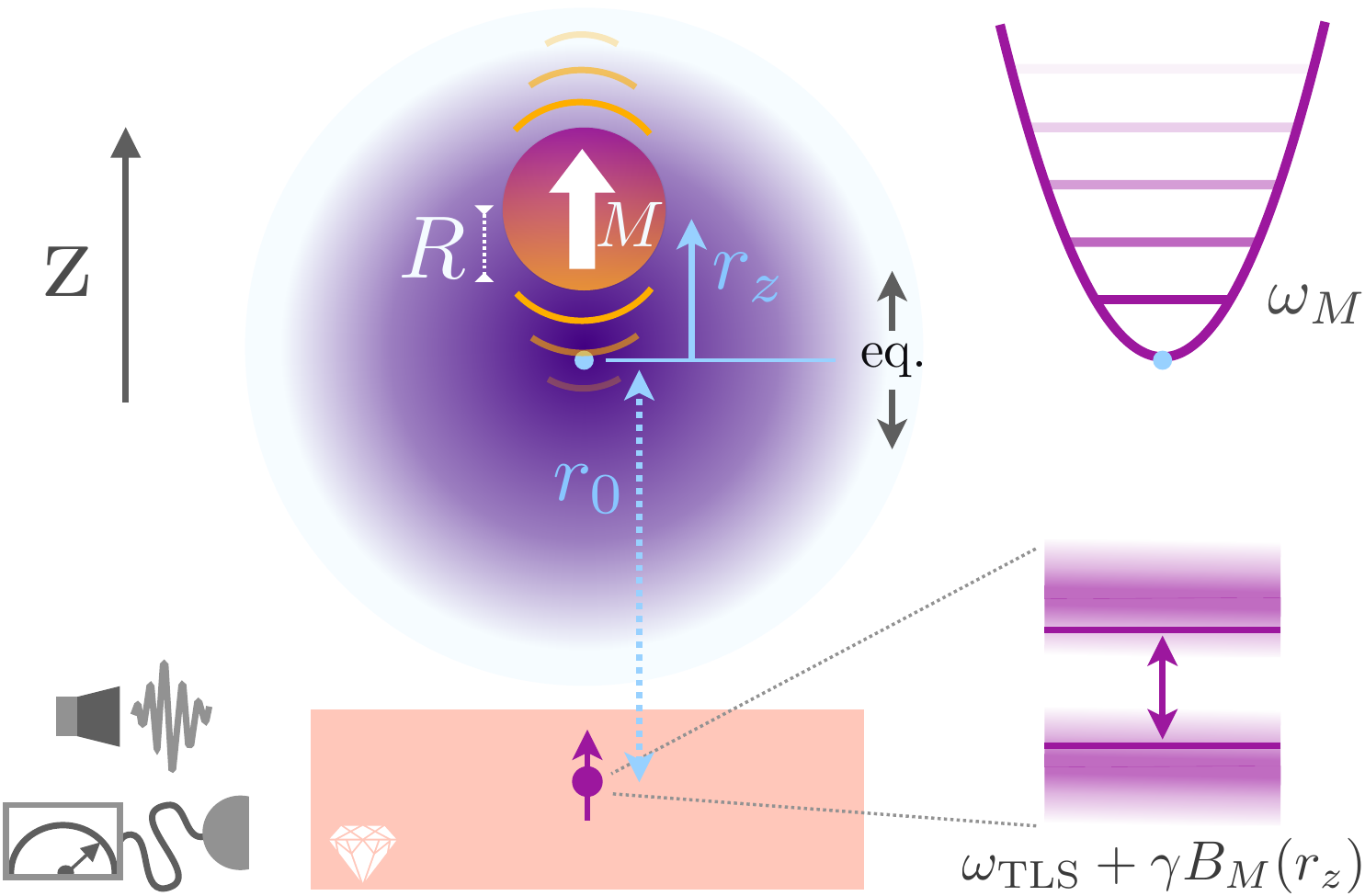}
	\caption{\label{fig:setup} {\bf Setup sketch.} A levitated magnetic particle of radius $R$ and magnetization $M$ follows a harmonic motion in the $Z$ direction with frequency $\omega_{M}$. A magnetically sensitive TLS (e.g. a NV center in diamond) of bare energy splitting $\omega_{\rm TLS}$ is placed in the axis of oscillation at a distance $r_0$ from the center of the harmonic potential. The energy splitting of the TLS is a function of the position of the magnet $r_z$. The position of the particle is initially described by a thermal distribution (indicated as the purple blurred area). The equilibrium position of the trap can be shifted.}
\end{figure}

In this Letter, we introduce a feedback cooling scheme that is capable of reaching the GS in low-frequency traps. Our protocol uses the TLS as a sensor that determines the position and momentum of the particle with an uncertainty comparable to the width of a coherent state, thereby transforming an initially mixed thermal state into a pure coherent state via a sequence of measurements. This allows the subsequent extraction of energy via a coherent displacement operation, such as a shift of the trap center to the position of the coherent state. The viability of our scheme becomes apparent when considering that the entropy of the initial thermal state is $S \approx\log_2(\bar{n}_{T=300 {\rm\, K}}) \approx 33$ bits, which indicates that, under ideal conditions, a pure state can be reached after just $\sim 33$ measurements with a binary sensor. This is in stark contrast to the $10^{10}$ measurements required for the previously mentioned proposals. To achieve this reduction in entropy within a minimal time and number of measurements, we designed an adaptive sensing scheme that ensures that the maximal amount of information is gained in each measurement. We numerically show the robustness of our method to imperfect readout fidelities and realistic heating rates.  Our adaptive sensing scheme is directly applicable to dc field sensing tasks and differs from previous schemes~\cite{cappellaro2012, dinani2019, scerri2020} in its focus on maximal information gain in each measurement and the possibility to mitigate backaction effects. The use of a TLS sensor also distinguishes our protocol from existing feedback schemes for optical setups, which operate in the continuous measurement regime with infinitesimal information gain and simultaneous feedback~\cite{magrini2020, tebbenjohanns2020a, walker2019, doherty1999a} as well as in the pulsed regime~\cite{vanner2011}. Moreover, the presence of a TLS provides a nonlinearity that can be employed in subsequent stages of the experiment for nonclassical state preparation or advanced sensing protocols~\cite{yin2013, scala2013, pedernales2020, chang2010, martinetz2020}.

{\it Setup and protocol.---}We consider a spherical magnetic particle of radius $R$, mass $m$ and magnetization $M$ levitated in vacuum, such that the dynamics of its c.m. is well approximated by three uncoupled harmonic oscillators. We assume that the magnetic dipole moment is aligned with the $Z$ direction and that a magnetically sensitive TLS, with bare energy splitting $\omega_{\rm TLS}$, is also aligned in the same axis at a distance $r_0$ from the equilibrium position of the trap; see Fig. \ref{fig:setup}. The magnetic moment generates a field at the position of the TLS that can be expanded to first order in terms of the displacement of the magnet from its equilibrium position, $r_z$, as  $B_M(r_z) \approx B_0 + G r_z$, provided that the position variance of the magnet $\Delta r_z$ fulfills the condition $\Delta r_z\ll r_0$. Here, $B_0 = 2 \mu_0 M R^3/(3 r_0^3)$, with $\mu_0$ the vacuum permeability, and ${G = - 2 \mu_0 M R^3/r_0^4}$. While the $B_0$ leads to an overall detuning of the TLS, $G$ provides a coupling $g = \gamma G a_0$ between the TLS and the position of the oscillator, where $\gamma$ is the gyromagnetic ratio of the TLS and $a_0 = \sqrt{\hbar/(m \omega_M)}$. We require the ultrastrong coupling regime $g \gg \omega_{M}$, with $\omega_M$ the trap frequency, which allows us to run our protocol in timescales where the harmonic dynamics of the oscillator are negligible and the system is well described by the Hamiltonian
\begin{equation}
\hat{H} = \hbar g \hat{z} \frac{\hat{\sigma}_{z}}{2} + \hbar \Omega(t) \frac{\hat{\sigma}_{x}}{2},
\label{eq:full-hamiltonian}
\end{equation}
which is stated in the rotating frame of the TLS. The last term represents a driving on the TLS, with time-dependent Rabi frequency $\Omega(t)$ and $\hat z = \hat r_z / a_0$ denotes the dimensionless position operator. We show in Supplemental Material~\cite{hc-supplementary-material} that the coupling to the particle motion in the $X$ and $Y$ directions can be neglected.

\begin{figure}
	\centering
	\includegraphics[width=\columnwidth]{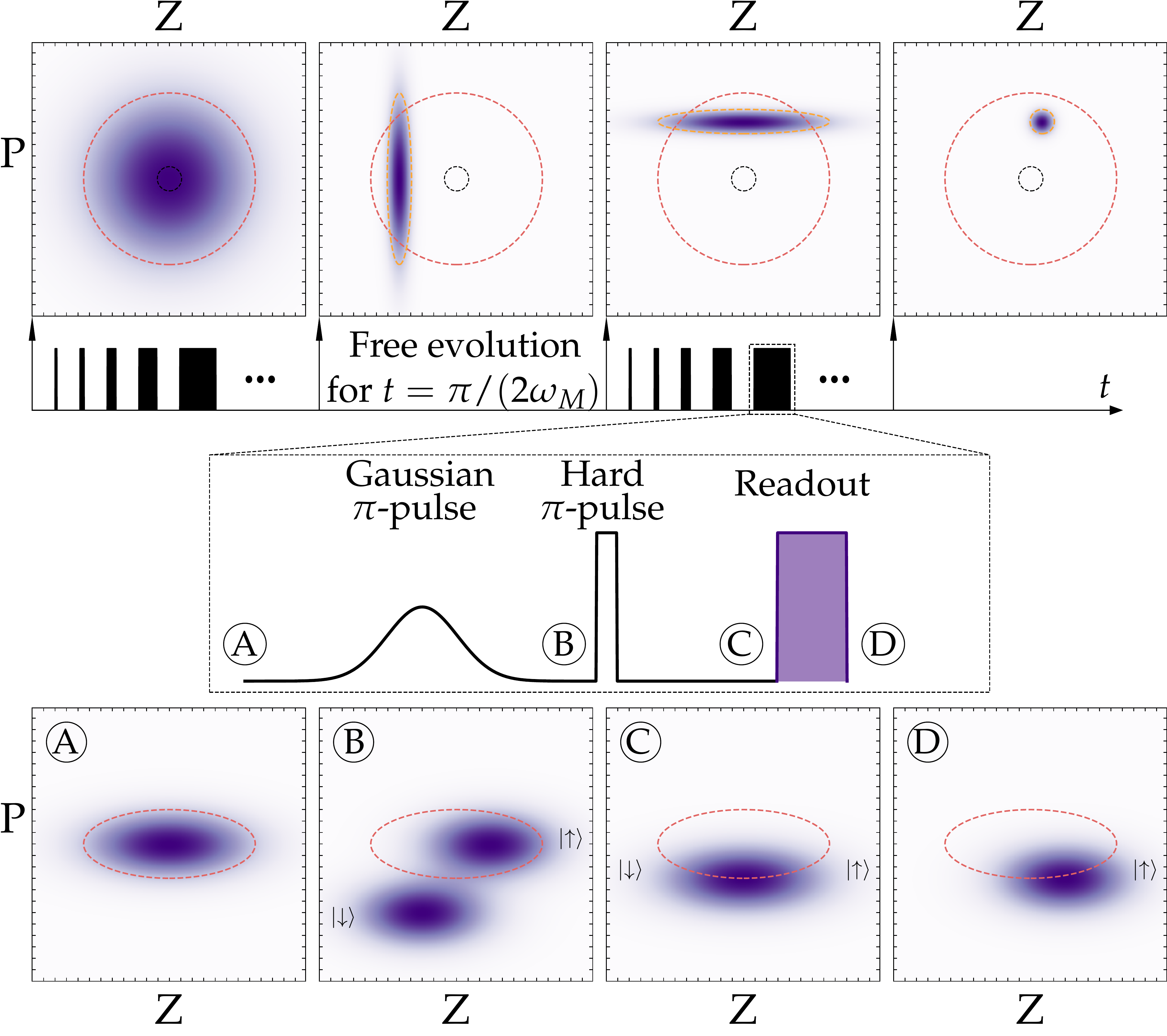}
	\raggedleft 
	\caption{{\bf Cooling scheme.} The upper panels show the Wigner function of the particle from its initial state till the end of the second part of the protocol. The middle section shows the amplitude profile of the pulse in each step. The lower section shows the evolution of the Wigner function during a single step of the protocol and highlights the displacement that is induced by the pulse.}
	\label{fig:protocol}
\end{figure}

Our protocol consists of three main parts: First, (i) we reduce the position uncertainty of the oscillator to that of the GS, and then (ii) we allow for a quarter of a rotation in phase space, which maps the momentum quadrature onto the position quadrature. Then we repeat the first step, which leaves the system in a coherent state. The protocol up to this point is visualized in the four upper panels of Fig. \ref{fig:protocol}. Finally, (iii) we displace the center of the trap to the position of the coherent state, cooling the system to the GS. We achieve the reduction of position uncertainty in parts (i) and (ii) by performing a sequence of adaptive measurements on the TLS. 

Each step starts with the TLS in its GS $\ket{\downarrow}$ and the oscillator in the mixed state $\rho_{M}$. The operator $\hat z$ introduces a shift of the bare TLS frequency, which allows us to correlate the TLS state with the particle position by performing a $\pi$ pulse that inverts the TLS state only for a specific range of detunings. A subsequent measurement of the TLS provides information about the particle position. To compute the spatial probability distribution associated with the measurement outcome $i = \{\uparrow, \downarrow\}$, we note that the Hamiltonian is diagonal in the spatial basis, which allows us to express the total evolution operator as $\hat U = \int \hat U'(z) \ketbra{z} \textrm{d}z$. The spatial probability distribution then follows by applying the projection operator $\mathcal{\hat P} = \ketbra{z} \otimes \ketbra{i}$ to the evolved state and taking the trace. The resulting expression is the familiar Bayesian update rule
\begin{align}
P_{n+1}\left(z \given i \right) {}&= \frac{1}{p_{n}(i)}\bra{z}\bra{i} \hat{U}_n \ket{\downarrow}\bra{\downarrow} \otimes \rho_{n} \hat{U}_n^\dagger \ket{i}\ket{z} \notag\\
&= \frac{I_n\left(z \given  i \right)}{p_{n}(i)} P_n\left(z\right).
\label{eq:recursion}
\end{align}
Here, $n$ denotes the measurement step, $p_{n}(i)$ is the probability of outcome $i$, and $I_n\left(z \given  \uparrow \right) = | \bra{\uparrow} \hat{U}'_n(z) \ket{\downarrow}|^2 = 1 - I_n\left(z \given  \downarrow \right)$ is the inversion profile of the pulse. The latter is a function of the eigenvalue of the dimensionless position operator $\hat z$, and its specific shape is determined by the pulse amplitude modulation $\Omega(t)$. We use a Gaussian form for $\Omega(t)$ which results in a Gaussian shape for $I_n\left(z \given  \uparrow \right)$. In Supplemental Material~\cite{hc-supplementary-material}, we describe the heuristic that we use to adaptively determine the mean and width of the Gaussian form of $\Omega(t)$ based on the current knowledge about the particle position which is encoded in $P_{n}\left(z\right)$. Our heuristic ensures that the width of $P_{n}\left(z\right)$ is reduced for both measurement outcomes and that the shape of the probability distribution remains close to a Gaussian. After each measurement Eq.~(\ref{eq:recursion}) is used to update the probability distribution. This can be done on a real-time computer with minimal resources, because the analytical form of $I_n\left(z \given  i \right)$ is known.

Our protocol also accounts for the measurement backaction on the oscillator. For perfect readout fidelities and Gaussian-shaped $\pi$ pulses, we find that the backaction amounts to a displacement of the momentum probability distribution for the $\ket{\downarrow}$ outcome, while no backaction occurs for $\ket{\uparrow}$. This is shown in the lower panels of Fig.~\ref{fig:protocol}. The displacement depends only on the pulse duration and is, therefore, known for every pulse~\cite{hc-supplementary-material}. Although it does not affect the spatial probability distribution, it must be kept track of, as it becomes relevant in the part of the protocol where the momentum quadrature is mapped onto the position quadrature. The fact that we are measuring a quantum system that can suffer backaction effects precludes an application of existing Ramsey-based adaptive sensing schemes~\cite{cappellaro2012, dinani2019, scerri2020}, because their backaction inevitably leads to a broadening of the momentum distribution~\cite{rao2016}.

If the TLS readout is imperfect, with fidelity $f < 1$, the measurement outcome loses its one-to-one correspondence to the TLS state, which is instead defined via the conditional probability distribution $F\left( o \given \uparrow \right)$, with $o = \{0,1\}$ denoting the outcome. In consequence, the spatial probability distribution after the measurement contains contributions from the oscillator state associated with TLS-state up and TLS-state down. This leads, assuming equal readout fidelities for both TLS states $F\left( o \given \uparrow \right)=F\left( o \given \downarrow \right)$, to a modified update rule~\cite{hc-supplementary-material}
\begin{equation}
	P_{n+1}\left(z \given 1 \right) = \frac{1}{p_n \left(1\right)}\left[f I_n\left(z \given  \uparrow \right)   + (1-f)I_n(z \given \downarrow) \right] P_{n}\left(z \right),
	\label{eq:recursion-f}
\end{equation}
which is stated here for outcome $o = 1$ and which has an analogous form for $o=0$. Such measurements still lead to a narrowing of the distribution in the position quadrature, but they introduce a broadening in the momentum quadrature, because the displacement is different for each of the two outcomes. To prevent this we introduce a hard $\pi$ pulse, which inverts the TLS state independently of its detuning. This swaps the TLS state that is associated with each of the two parts of the oscillator state, and a subsequent free evolution leads to a displacement of both parts. The final displacement of both parts of the oscillator state is nonzero but equal, preventing the broadening of the momentum distribution and making this case equivalent to that of $f=1$. The corresponding sequence is shown in Fig. \ref{fig:protocol}.

\begin{figure}[t!]
	\includegraphics[width=\columnwidth]{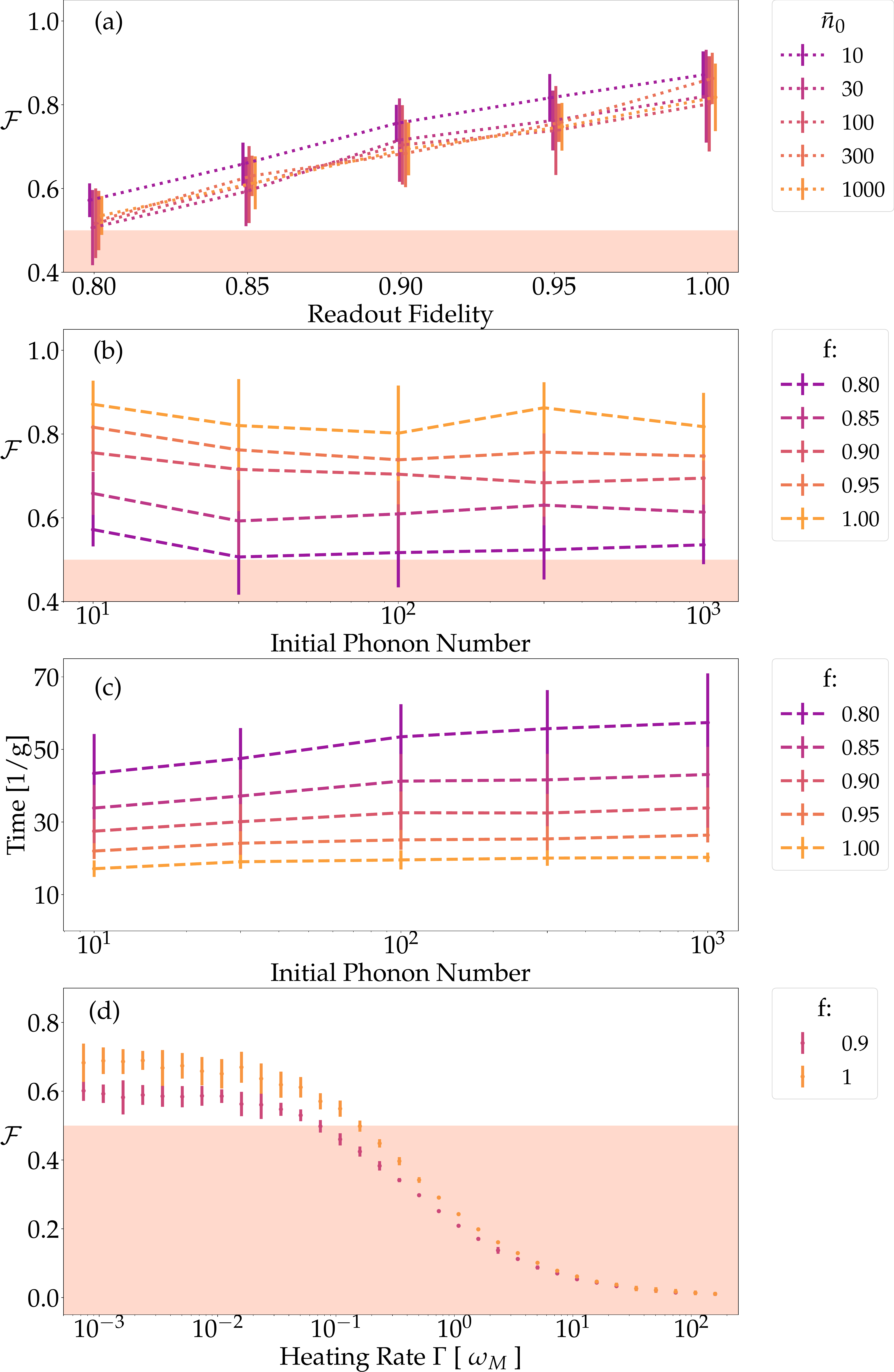}
	\caption{{\bf Numerical simulations.} (a) GS fidelity $\mathcal{F}$ at the end of the protocol versus readout fidelity for different initial phonon numbers. (b) The Same data as in (a) plotted as a function of the initial phonon number. These results were obtained by simulating the von Neumann equation of the closed system. (c) Time needed for part one of the protocol in units of inverse coupling $1/g$. (d) GS fidelity $\mathcal{F}$ as a function of the heating rate for different readout fidelities and an initial state with $\bar{n} = 100$ that is in thermal equilibrium with the environment. The data were generated by simulating the Lindblad equation with $\omega_{M} = 2 \pi g / 300$. The shaded areas mark the region with GS fidelity below 0.5, which for a thermal state corresponds to a mean phonon number larger than one. The GS fidelity of state $\rho$ is $\mathcal{F}=\bra{0}\rho\ket{0}$~\cite{hc-supplementary-material}.}
	\label{fig:results}
\end{figure}

{\it Numerical simulations.---}To demonstrate the performance of our protocol, we numerically simulate the full quantum dynamics of a thermal initial state~\cite{hc-supplementary-material}. We use the GS fidelity as a figure of merit instead of the mean phonon number, because, due to the properties of our protocol, any deviations from the GS are of a nonthermal nature. Hence, GS fidelities are significantly higher than those of a thermal state with the same mean phonon number. Figure~\ref{fig:results}(a) shows that high GS fidelities can be reached even for readout fidelities significantly below unity and high initial phonon numbers. In fact, as our protocol makes no assumption on the initial phonon number and Fig.~\ref{fig:results}()b) shows no deterioration of the performance with increasing initial thermal occupancies, we expect a similar performance at room temperature, for which numerical simulations become infeasible. This is a key distinguishing factor to previous TLS-based cooling proposals~\cite{rabl2009, lau2016, romero-isart2012, puebla2020, jaehne2008}. 

The time needed by our protocol to reach the minimal width in each quadrature is dominated by the last few measurements. It is, therefore, almost independent of the initial phonon number as shown in Fig.~\ref{fig:results}()c). This property follows from the fact that the duration of each step is inversely proportional to the width of the spatial distribution in that step. Notice that we are neglecting the readout time, as this is highly dependent on the experimental implementation. For readout fidelities of 0.9 and moderate initial phonon numbers of $\bar{n}\approx 10^5$, the number of measurements for each quadrature can be kept below 75. Moreover, the total protocol duration is set by the trap frequency, which, in turn, determines the cooling rate. Figure~\ref{fig:results}(d) shows that GS fidelities of 0.5 can be reached for heating rates on the order of $\Gamma \approx \omega_{M}/10$. In future work, the update rules in Eq.~(\ref{eq:recursion}) and (\ref{eq:recursion-f}) can be modified to account for the effect of heating on the probability distributions, which should allow higher heating rates.

{\it Experimental feasibility.---}Our protocol is capable of reaching the GS under the assumption that the coupling between the TLS and the particle, $g$, fulfills two conditions: (i) It is larger than the inverse of the TLS coherence time, $g > 1/T_2$, and (ii) it is in the ultrastrong coupling regime $g \gg \omega_M$. Additionally, numerical simulations show that the protocol performs best when (iii) the readout fidelity of the TLS is above 0.8 (in order to reach GS fidelities above 0.5) and (iv) the motional heating rate $\Gamma$ is lower than the trap frequency, $\Gamma  < \omega_{M}$. We analyze the experimental feasibility of these requirements by considering a particle with a density of $7 \times 10^3 \textrm{ kg/m}^3$~\cite{gieseler2020} and a TLS implementation based on a single nitrogen-vacancy (NV) center~\cite{wu2016} at two different implantation depths: shallow (case {\bf A}) and deep (case {\bf B}). The specific parameters are provided in Table~\ref{tab:nv-parameters}. Our protocol requires coherence times of $10$ and $200$ $\mu$s for cases {\bf A} and {\bf B}, respectively, which is well within reported values. For shallow NV centers, room-temperature coherence times as long as $250\ \mu$s have been demonstrated~\cite{staudacher2012}, as well as fast single-shot readout at cryogenic temperatures with fidelities of $78.6 \pm 2.5\%$~\cite{irber2021}. For a scenario like that of case {\bf{B}}, coherence times of up to $2.4$ ms~\cite{herbschleb2019} and single-shot readout fidelities over 92\%  at cryogenic temperatures~\cite{robledo2011} have been reported. A room-temperature implementation of our protocol with NV centers would also be possible, albeit the high and precisely aligned external magnetic fields, which are required to achieve high-fidelity readout through a mapping of the NV state onto the nuclear spin~\cite{neumann2010}, represent an added experimental challenge. These estimates indicate that our protocol can be used for the GS cooling of particles of $1-10\ \mu$m radius, in contrast to reported GS cooling experiments~\cite{delic2020, magrini2020} with particles one order of magnitude smaller.

The final step of our protocol, the shift of the trap center, can be performed with a current loop that introduces an additional magnetic field in analogy to the feedback mechanisms in optical setups~\cite{magrini2020}. Such a shift can even be implemented with a piezoactuator as, due to the low trap frequency, the involved timescales are slow.

\begin{table}
	\raggedright
	\begin{tabular}{ p{1.5cm}  p{1.5cm}  p{1.5cm}  p{1.5cm}  p{1cm}  p{1cm}}
		\hline
		\hline
		Case & {\centering $R$} & $\omega_{M}/(2\pi)$ & $d$ & $d_{\rm NV}$ & $g$  \\ [0.5ex]
		\hline
		\bf{A} & 0.5 & 1 	& 0.65 & 0.1 & 148  \\ [0.5ex]
		\bf{B} & 5 	 & 0.1 	& 7    & 1   & 6 \\ [0.5ex]
		\hline
	\end{tabular}
	\caption{{\bf Two parameter regimes with NV centers.} $R$ is the particle radius, $\omega_{\rm M}$ the trap frequency, $d$ the distance between particle center and NV, $d_{\rm NV}$ the implantation depth of the NV center, and $g$ the coupling between the magnet and the NV. Distances are stated in microns and frequencies in kilohertz.}
	\label{tab:nv-parameters} 
\end{table}

Our protocol is well suited for the experimental setup in Ref.~\cite{gieseler2020}, which can operate in the ultrastrong coupling regime and where the authors expect to reach couplings of $2.6$~kHz and heating rates of $0.8$~kHz. Although existing dissipative TLS cooling schemes~\cite{rabl2009, lau2016, romero-isart2012, puebla2020, jaehne2008} could also be applied in such a setup, they require an initial phonon number approximately $10-100$ times lower than the quality factor of the resonator~\cite{rabl2010}, which necessitates additional precooling mechanisms. Remarkably, our protocol does not have such a limitation, which makes it applicable even in room-temperature environments, if sufficiently high readout fidelities are available. While the heating rate $\Gamma = k_b T / (\hbar Q)$ increases with the temperature of the environment, leading to higher final phonon numbers, the cooling mechanism does not completely break down, unlike in the previously mentioned schemes.
\\

\textit{Conclusion.---}By splitting the process of cooling into two steps: entropy reduction and energy extraction our protocol is able to overcome the limitations present in dissipative TLS-based cooling schemes when operating at low-frequencies. Thus, it constitutes, to the best of our knowledge, the first viable GS-cooling scheme for micron-scale particles in low frequency traps.  In particular, an implementation of our ideas using NV centers in cryogenic environments can reach the required parameter regime.  Moreover, the ideas presented here can be independently applied as an adaptive sensing scheme, suitable for conditions where the prior information on the unknown parameter is limited, or where the backaction on the sensed system becomes relevant~\cite{scerri2020}. Our protocol can be improved further by a rigorous optimization of the parameters of the adaptive algorithm and by extending it to multiple sensors. More involved TLS drivings can be used to lift the requirement of the ultrastrong coupling regime and to allow for even higher heating rates.

\begin{acknowledgments}
{\it Acknowledgments.---}We thank J.F. Haase, B. D'Anjou, M. Korzeczek and P. Schmidt for helpful comments on the manuscript. We acknowledge support by the ERC Synergy grant HyperQ (Grant No. 856432), the EU projects HYPERDIAMOND (Grant No. 667192) and AsteriQs (Grant No. 820394), the QuantERA project NanoSpin (13N14811), the BMBF project DiaPol (13GW 0281C), the state of Baden-W\"urttemberg through bwHPC, the German Research Foundation (DFG) through Grant No. INST 40/467-1 FUGG, and the Alexander von Humboldt Foundation through a postdoctoral fellowship.
\end{acknowledgments}

\clearpage
\onecolumngrid
\setcounter{figure}{0}
\renewcommand\thefigure{S\arabic{figure}}  
\setcounter{equation}{0}
\renewcommand{\theequation}{S\arabic{equation}}
\renewcommand*{\HyperDestNameFilter}[1]{\jobname-#1}

\section{\Large Supplemental Material}
\vspace{2\baselineskip}

\section{Coupling to Additional Modes}
\label{sup:add-modes}
In typical experimental arrangements that are relevant for our cooling scheme \cite{gieseler2020-2} the particle is trapped in all three directions with similar frequencies. However, we have neglected the coupling to the X and Y-modes in the main text. Here we show that the contribution to the signal from these modes is indeed negligible.

We start with the dipole field created by the particle
\begin{equation}
	\mathbf{B}(\mathbf{r}) = \frac{\mu_0}{4 \pi} \frac{3 \mathbf{r} (\mathbf{m} \cdot \mathbf{r})}{r^5} - \frac{\mathbf{m}}{r^3}.
\end{equation}
The X and Y components of this vector field can be neglected because they couple to the $\sigma_x$ and $\sigma_y$ operators of the TLS which rotate with the TLS eigenfrequency. Therefore these contributions average to zero over the relevant timescales. The coupling to the X and Y-modes is given by the derivatives of the Z-component along x and y, evaluated at the trap center $x=0$, $y=0$
\begin{align}
	\frac{\partial B_z}{\partial x} \biggr\rvert_{x=0, y=0} &= \frac{\mu_0}{4 \pi} \frac{3 m_x z}{z^5}, \\
	\frac{\partial B_z}{\partial y} \biggr\rvert_{x=0, y=0} &= \frac{\mu_0}{4 \pi} \frac{3 m_y z}{z^5}.
\end{align}
We see that the couplings are proportional to the transverse components of the magnetization and therefore on the rotation of the particle. 

Assuming that it is possible to set the preferential direction for the trap along the Z-axis, we estimate the maximal rotation due to thermal fluctuations. The amplitude of this rotation can be estimated by setting the change in potential energy equal to the thermal energy
\begin{equation}
	\Delta U(\Delta \theta) = k_b T,
\end{equation}
where $\Delta \theta$ is the polar angle enclosed between the Z-axis and the magnetization vector of the particle. By using the derivation in the Supplemental Material of \cite{gieseler2020-2} we obtain a maximal rotation of $\Delta \theta_{\textrm max} = 2 \cdot 10^{-5}$ for a particle with radius $R = 1 \mu$m and a levitation height of two times the radius at a temperature of $T=4$ K.

Assuming a frequency of the X and Y-modes \cite{gieseler2020-2} on the order of 100 Hz we obtain thermal populations of the modes on the order of $\bar{n} \approx 10^8$. Therefore, the maximal contribution to the detuning of the TLS due to the thermal motion of the particle along the X and Y-directions is given by
\begin{equation}
	\frac{\Delta_{x,y}}{g_z} = \frac{g_{x,y}}{g_z} \sqrt{\bar{n} + \frac{1}{2}} = \frac{2 \cdot 10^{-5}}{2} \sqrt{\bar{n} + \frac{1}{2}} \approx 0.1,
\end{equation}
where $g_{x,y,z}$ denote the couplings to the respective modes. We see that even a thermal state in the X or Y-mode leads to a contribution one order of magnitude smaller than a coherent state in the Z-mode, which produces a detuning of $\Delta_z = g_z$. We can therefore conclude that the contributions of the X and Y-modes can be neglected.

\section{Shaped Inversion Profiles}
\label{sec:ShapedInversionProfiles}
In this section, we present a detailed derivation of the Bayesian update rule in Eq. (2) of the main text, elucidate our choice of the Gaussian inversion profile, describe the modulation of the drive amplitude that is needed to generate this profile and explain the properties of such a pulse, including the backaction that it induces on the particle. The Hamiltonian in Eq. (1) of the main text
\begin{equation}
	\hat{H} = \hbar g \hat{z} \frac{\hat{\sigma}_{\rm z}}{2} + \hbar \Omega(t) \frac{\hat{\sigma}_{\rm x}}{2},
	\label{eq:sl-full-hamiltonian}
\end{equation}
is the starting point of our analysis. It is diagonal in the spatial basis, allowing us to express the evolution operator as $\hat{U} = \int \hat{U}'(z) \ket{z}\bra{z} \textrm{d}z$, where $U'(z)$ acts in the TLS subspace and is the unitary evolution operator associated to Hamiltonian~(\ref{eq:sl-full-hamiltonian}) after substituting the dimensionless position operator $\hat z$ by the eigenvalue $z$ of the corresponding eigensate $\ket{z}$. The goal of our protocol is to reduce the widths of the spatial and momentum probability distributions to the width of a coherent state. The effect of a single evolution and measurement step on the spatial probability distribution can be easily computed for each measurement outcome $i= \{ \uparrow, \downarrow \}$
\begin{equation}
	\begin{split}
		P_{n+1}\left(z \given i \right) &= \frac{1}{p_{n}(i)}\bra{z}\bra{i} \hat{U}_n \ket{\downarrow}\bra{\downarrow} \otimes \rho_{n} \hat{U}_n^\dagger \ket{i}\ket{z} \\
		&=\frac{| \bra{i} \hat{U}'_n(z) \ket{\downarrow}|^2}{p_{n}\left(i\right) } P_n\left(z\right)  = \frac{I_n\left(z \given  i \right)}{p_{n}(i)} P_n\left(z\right).
		\label{eq:sl-recursion}
	\end{split}
\end{equation}
The index $n$ indicates the measurement number, $p_{n}(i)$ the probability of outcome i, and $P_n(z) = \bra{\downarrow}\bra{z} \rho_n \ket{z}\ket{\downarrow}$ the spatial probability distribution at the beginning of the step. The initial probability distribution $P_0(z)$ corresponds to that of a thermal state and is known before the start of the experiment. To compute the post-measurement distribution only the knowledge of the inversion profile $I_n\left(z \given  \uparrow \right) = |\bra{\uparrow} \hat{U}'_n(z) \ket{\downarrow}|^2 \le 1$, which corresponds to the transition probability as a function of $z$, is required. This depends on the specific form of the amplitude modulation $\Omega (t)$ and has no general analytical solution. Nevertheless, to avoid its computation in each step of the protocol, its functional form can be precomputed and fitted to a parametrized curve, such that this analytic approximation can be used in the update rule~(\ref{eq:sl-recursion}) during the experiment. Provided that this parametrized curve fits the real inversion profile well, this will not corrupt the functioning of our protocol. Thus, to obtain the probability distribution after each measurement, the approximated inversion profile simply has to be multiplied with the previous probability distribution. Of course, this procedure becomes exact when the analytical solution of the inversion profile is known. For example, consider a pulse with constant Rabi frequency $\Omega (t) \equiv \Omega$ and duration $\tau$, such that $\int_0^\tau \Omega \textrm{d}t = \pi$, i.e. a square $\pi$-pulse; the analytical form of the inversion profile is
\begin{equation}
	I\left(z \given  \uparrow \right) = \frac{\Omega^2}{\Omega^2 + \Delta(z)^2} \sin\left( \frac{\sqrt{\Omega^2 + \Delta(z)^2}}{\Omega} \frac{\pi}{2}\right)^2,
\end{equation}
which is parametrized by the Rabi frequency $\Omega$ and where $\Delta(z) = g z$. Even though square pulses are not used in our protocol we note that the width of this inversion profile is dependent on the amplitude $\Omega$ and, therefore, the duration of the pulse, which is a common feature of all pulses.

\begin{figure}
	\includegraphics[width=\textwidth]{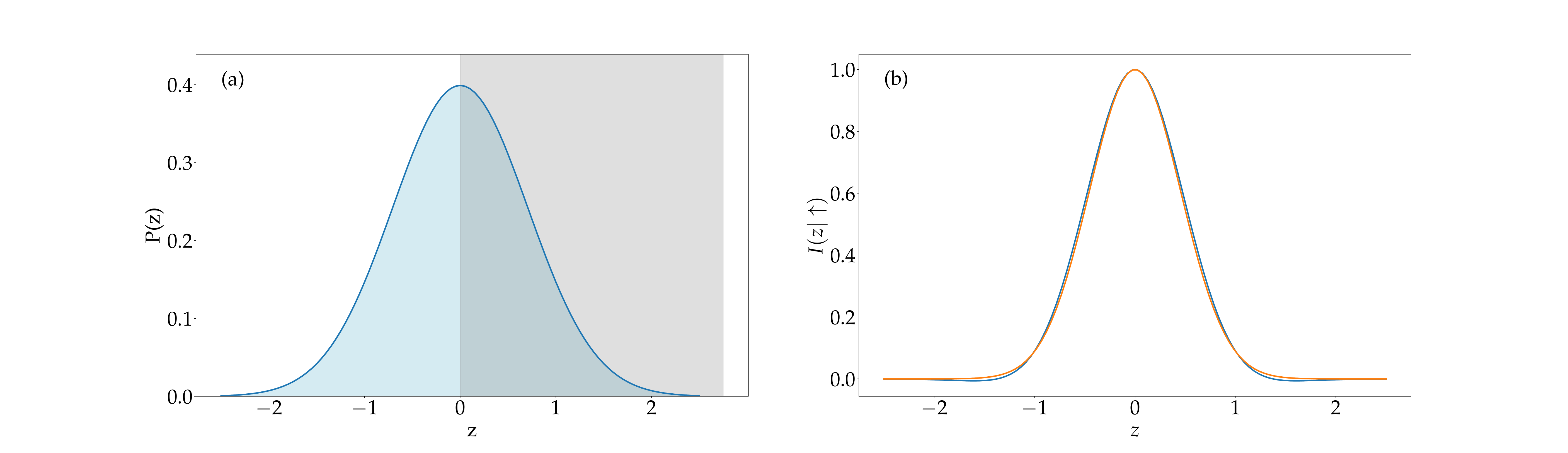}
	\caption{(a) The Gaussian prior distribution is shown in blue and the shaded area indicates the part covered by the rectangular inversion profile. They are aligned such that the inversion profile covers half of the distribution. (b) Comparison of the inversion profile generated by a Gaussian modulation (blue) of the driving amplitude and a Gaussian function with the parameters stated in the text (orange).}
	\label{fig:sl-rectangular-profile}	
\end{figure}

One of the most important requirements when choosing the appropriate inversion profile is that the measurement must yield a significant reduction in entropy. Furthermore, this reduction must be achievable not only for the initial distribution, but also for all intermediate distributions. This requirement is fulfilled for a rectangular inversion profile that correlates half of the distribution with the TLS-state up and the other half with TLS-state down as shown in Fig.~\ref{fig:sl-rectangular-profile}(a). This means that the probabilities of measuring either TLS state are $1/2$ (in case of perfect readout) and that the posterior distribution for either outcome will have one bit of entropy less than the prior distribution. It is obvious that this halving of the distribution can be continued in the subsequent measurements. However, for readout fidelities $f<1$ such a pulse profile proved to be unstable as it lead to discontinuities in the probability distribution due to its rectangular (discontinuous) form. For our protocol, we choose to use a Gaussian inversion profile parameterized by the mean $\mu_{\rm I}$ and the variance $\sigma_{\rm I}$ as
\begin{equation}
	I\left(z \given  \uparrow \right) = \exp\left( - \frac{(z-\mu_{\rm I})^2}{2 \sigma_{\rm I}^2}\right).
\end{equation}	
Given that our initial (thermal state) and target (coherent state) distributions are Gaussian and that a product of two Gaussians (prior times inversion profile) yields another Gaussian, we expect that upon measuring outcome $\ket{\uparrow}$ the distribution will remain Gaussian. However, we expect slight deviations because the inversion profile for the $\ket{\downarrow}$ outcome is given by $I_n\left(z \given  \downarrow \right) = 1 - I_n\left(z \given  \uparrow \right)$, which is not Gaussian. It is shown in the next section that despite this fact the deviations are typically small. 

To the best of our knowledge, the inversion profile associated to a Gaussian modulation of the pulse amplitude $\Omega(t)$ has no analytic solution. However, it has been shown that it closely resembles a Gaussian profile~\cite{bauer1984}, which is sufficient for the proper functioning of our protocol. In particular, for a modulation of the form
\begin{equation}
	\Omega(t) = \frac{\pi}{\sqrt{2 \pi \sigma_{\rm p}^2}}\exp \left( -\frac{ (t-\tau/2)^2}{2 \sigma_{\rm p}^2} \right),
\end{equation}
the numerically computed inversion profile can be fitted to a Gaussian of variance $\sigma_I^2 = 1 / (2 g \alpha \sigma_p^2)$. Here, the factor of $2$ is a consequence of the square in the definition of the inversion profile, and the constant $\alpha$ will vary with the duration of the pulse $\tau$, and needs to be numerically determined. For the simulations in this work, we chose a pulse duration of $\tau = 10 \cdot \sigma_p$, for which we find that the value $\alpha \approx 1.15$ yields an  approximation of the inversion profile that is sufficiently good for the functioning of our protocol. A comparison between the numerically determined profile and the Gaussian approximation with the stated parameters is shown in Fig.~\ref{fig:sl-rectangular-profile}(b). We note that a too-short pulse duration leads to a windowing, effect which results in undesired oscillations on top of the Gaussian form. 

\begin{figure}
	\includegraphics[width=0.45\textwidth]{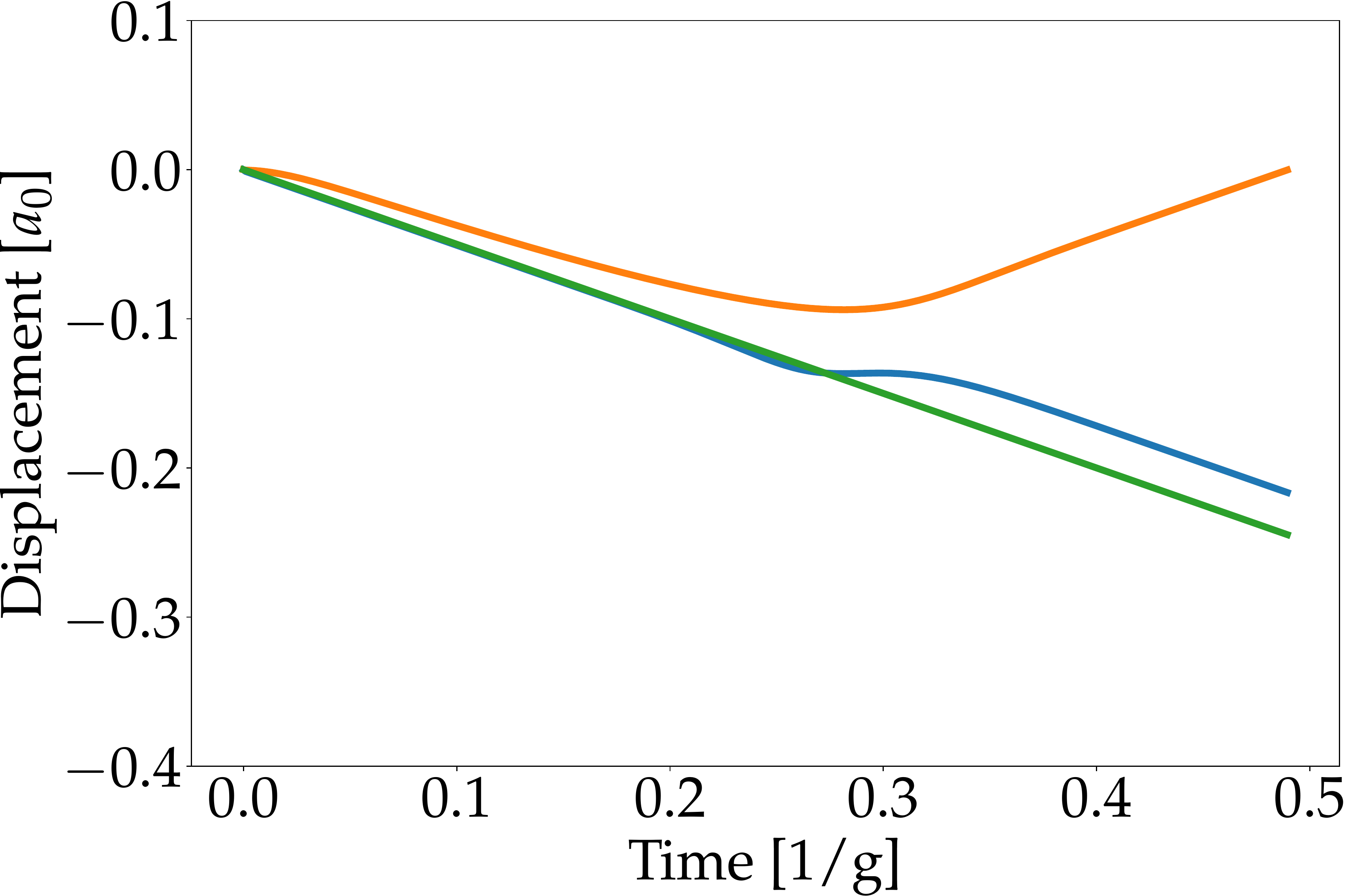}
	\caption{Evolution of the particle displacement during the Gaussian pulse that is associated with TLS-state down (blue), TLS-state up (orange) and the displacement that would be expected under the action of $H = -g \hat{z}/2$ (green).}
	\label{fig:si-displacement}	
\end{figure}

In addition to correlating the TLS and particle states, the pulse also introduces a displacement of the particle. The numerically determined displacement of the particle is shown in Fig. \ref{fig:si-displacement}. The displacement of the particle state that is associated with the TLS-state up outcome is always zero while the state associated with TLS-state down gets displaced. Our simulations show that when the relationship between pulse duration $\tau$ and pulse width $\sigma_p$ is kept constant the displacement becomes linearly dependent on the pulse duration. Therefore the imposed displacement is always known to the experimenter without additional computational effort.

\section{Adaptive Sensing Algorithm for Gaussian Inversion Profiles}

A key ingredient in our protocol is the adaptive sensing scheme that ensures efficient entropy reduction in each measurement and a final distribution that is close to a Gaussian. In this section we describe how the pulse parameters are determined in each step of the algorithm based on the current spatial probability distribution. 

At the beginning of the algorithm we assume that the system is in a thermal state which means that the probability distributions for both quadratures are of the form
\begin{equation}
	P_0(z) = \frac{1}{2 \pi \sqrt{\bar{n} + \sigma_0^2}} \exp{\left( -\frac{z^2}{2 (\bar{n} + \sigma_0^2)} \right)},
\end{equation}
with $\sigma_0 = 1/\sqrt{2}$ denoting the width of a coherent state and 
\begin{equation}
	\bar{n} = \frac{e^{ -\frac{\hbar \omega_M}{k_B T}}}{1 - e^{ -\frac{\hbar \omega_M}{k_B T} }}
\end{equation}
the thermal occupation number, with $k_B$ denoting Boltzmann's constant and $T$ the temperature. The prior distribution is saved as a vector and updated in each step by applying the Bayesian update rule. Our algorithm is a Bayesian inference algorithm where $P_0(z)$ corresponds to the prior distribution and the inversion profile $I(z|\uparrow)$ to the likelihood function. As in other inference algorithms of this type, the precise form of the prior is not important for the proper functioning of the algorithm as long as it assigns a non-zero probability to the true outcome. Therefore, precise knowledge of $\bar{n}$ is not required. It is safe to assume a too-high initial temperature, because, due to the fact that the number of measurements grows logarithmically with $\bar{n}$, this will not significantly affect the efficiency of the protocol. However, a too-low $\bar{n}$ can lead to systematic errors, because the overlap of the true state with the prior distribution might become too small.

\begin{figure}
	\centering
	\includegraphics[width=\textwidth]{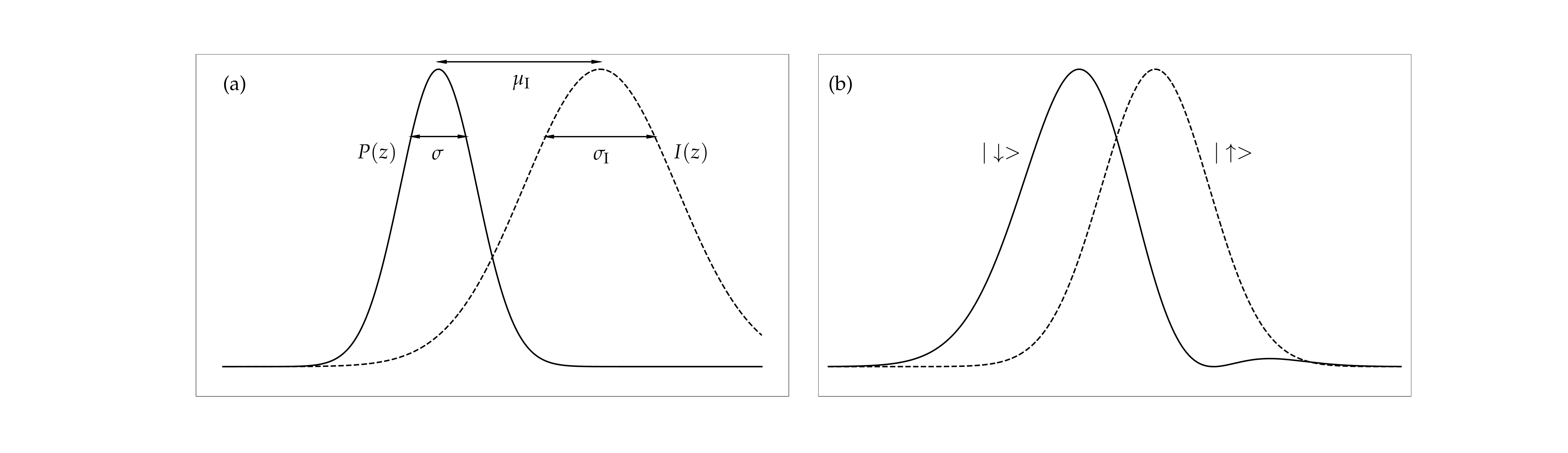}
	\caption{(a) Shows the parameters of the adaptive protocol. (b) Shows the posterior distributions for the two measurement outcomes. For this plot a Gaussian was used as the prior.}
	\label{fig:si-parameters}
\end{figure}

Given a Gaussian prior distribution centered at $\mu$ and with a variance $\sigma$, we need to determine the detuning $\mu_{\rm I}$ and the width $\sigma_I$ of the Gaussian inversion profile. These parameters are visualized in Fig. \ref{fig:si-parameters}(a). We choose the width to be linearly related to the width of the prior distribution via a multiplicative factor $w$ that is kept fixed throughout the protocol
\begin{equation}
	\sigma_I = w \cdot \sigma.
	\label{eq:si-update-width}
\end{equation}
Given $\sigma_{\rm I}$, we determine the detuning by fixing the probability of the TLS-state up outcome and inverting the relationship
\begin{equation}
	p_\uparrow = \int_{-\infty}^{\infty} \textrm{d}z \frac{1}{\sqrt{2 \pi \sigma^2}} e^{ - \frac{(z-\mu)^2}{2 \sigma^2} } e^{- \frac{(z-\mu_{\rm I})^2 }{2 \sigma_{\rm I}^2}}.
\end{equation}
This leads to
\begin{equation}
	\mu_{\rm I} = \mu \pm \sqrt{  - 2 (\sigma^2 + \sigma_{\rm I}^2) \log{\left( p_\uparrow \sqrt{ \frac{ \sigma^2 + \sigma_{\rm I}^2 }{\sigma_{\rm I}^2 }  } \right)} }.
	\label{eq:si-update-detuning}
\end{equation}
This parametrization was chosen because the outcome probability is a good measure of the equality of the entropy reduction over the two outcomes. For a given $\sigma_{\rm I}$, $p_\uparrow = p_\downarrow = 0.5$ leads to a more equal (but not exactly equal!) entropy reduction than $p_\uparrow < p_\downarrow$, which would lead to a higher entropy reduction for the spin up outcome. Hence, $\sigma_{\rm I}$ gives us a handle on how much entropy can be extracted in the measurement and $p_\uparrow$ determines how this reduction is distributed over the two possible outcomes. Note that both update rules in Eq. (\ref{eq:si-update-width}) and Eq. (\ref{eq:si-update-detuning}) assume that the prior distribution has a Gaussian form and require the knowledge of the width $\sigma$ of that Gaussian.

\begin{figure}
	\centering
	\includegraphics[width=\textwidth]{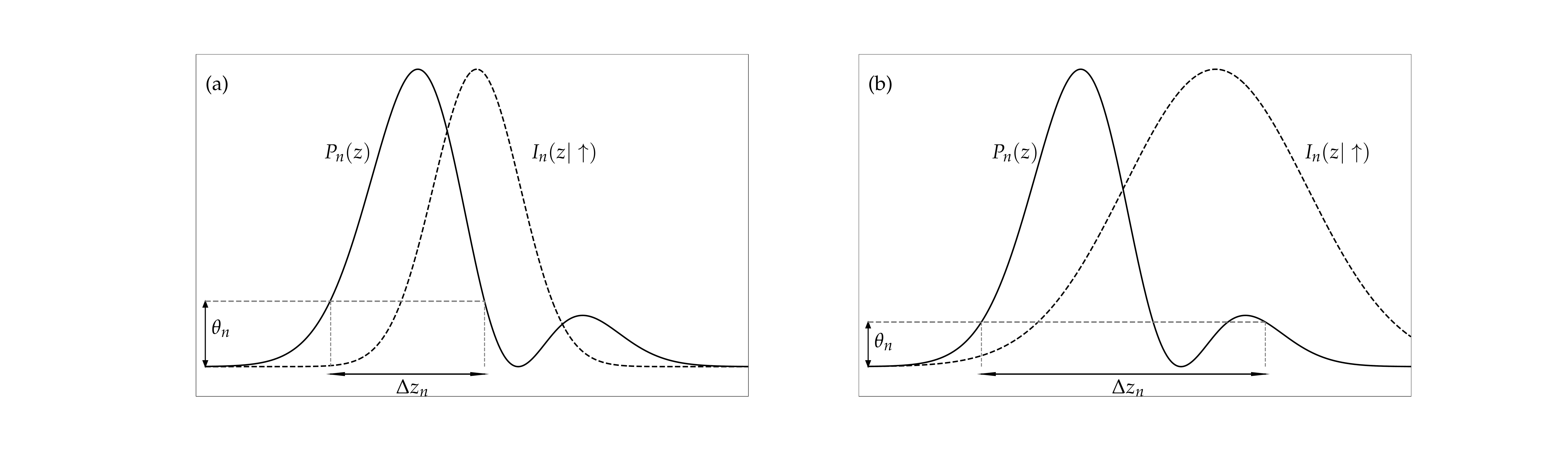}
	\caption{(a) The threshold $\theta_n$ is larger than the maximal point of the side peak. In this case the side peak will grow in the next measurement. (b) The threshold $\theta_n$ is smaller than the side peak leading to a wide inversion profile and therefore a suppression of the side peak in the next measurement.}
	\label{fig:si-side-peak}
\end{figure}

\begin{figure}
	\centering
	\includegraphics[width=0.75\textwidth]{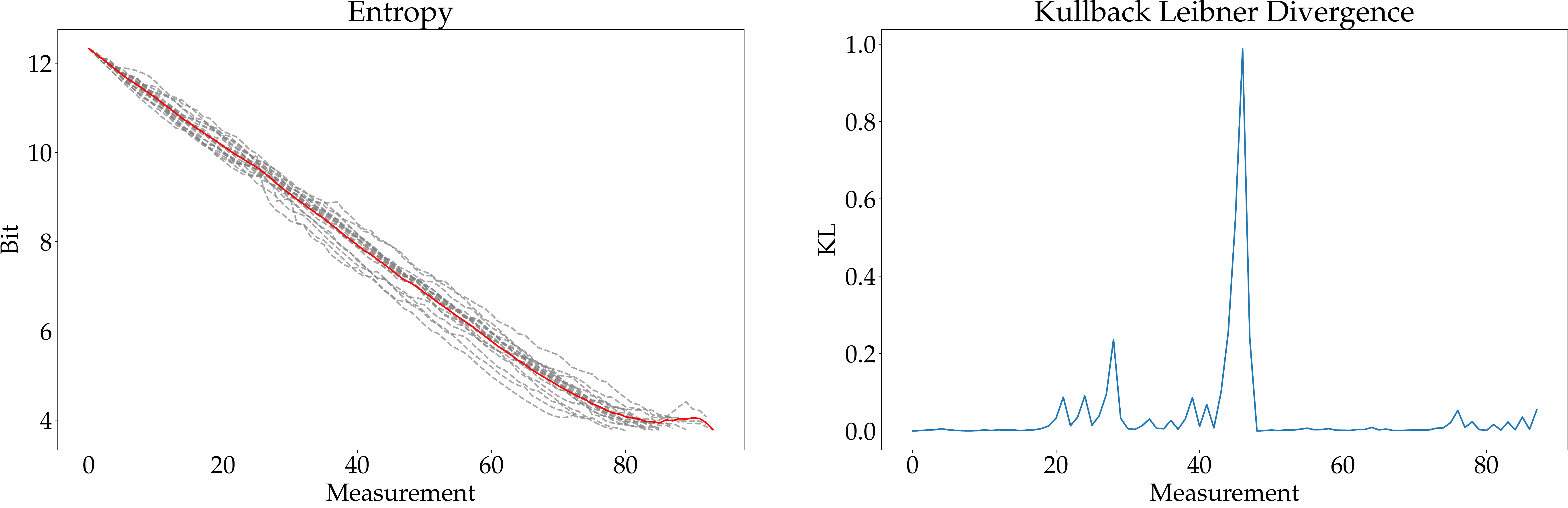}
	\caption{\raggedright (left) Entropy reduction in each measurement. The dashed lines show the trajectories of individual simulations, the red line shows the average over these trajectories. The parameters used for this plot are $\bar{n} = 300$, $f=0.9$. (right) Kullback-Leibner divergence for one of the trajectories in the left plot.}
	\label{fig:si-entropy-kl}
\end{figure}

To obtain an approximately deterministic cooling rate, the entropy reduction for both measurement outcomes should be equal. An alternative approach is to tune the algorithm parameters such that the entropy reduction averaged over the two outcomes (weighted by their probabilities) is maximal. However, this approach leads to multi-peaked distributions that violate our simplifying assumption that the distribution is close to a Gaussian in each step. As shown in Fig.~\ref{fig:si-parameters}(b), a small side peak always emerges for the spin down outcome. To ensure that this peak remains small, we choose the width of the inversion profile to be larger than the width of the prior distribution $\sigma_I > \sigma$ and the probability of detecting the TLS-state up to be $p_{\uparrow} < 0.5$. The prior condition makes the side peak wider and therefore lower, but also reduces the amount of entropy that can be extracted in the measurement. The latter condition decreases the absolute size of the side peak and makes the entropy reduction more unequal over the measurement outcomes. With these two considerations it can be ensured that the distribution remains close to a Gaussian. The entropy reduction for the TLS-state down outcome gives a lower bound on the entropy that can be extracted with each measurement. This deterministic decrease in entropy is shown in Fig.~\ref{fig:si-entropy-kl}(a).

To design the inversion profile, we have assumed that, in every step $n$, the prior probability distribution is a Gaussian, which, as discussed in Sec.~\ref{sec:ShapedInversionProfiles}, will not be the case in general. Nevertheless, our algorithm will still work in this case, albeit with reduced efficiency. To that end, we need to establish a methodology to associate, in each step, a Gaussian width $\sigma_n$ to the actual probability distribution $P_n(z)$, to be used in the update formulas~(\ref{eq:si-update-width}-\ref{eq:si-update-detuning}). More importantly, we want to do this in a way that avoids the deviations from accumulating as the protocol progresses. With that in mind, we define a threshold value $\theta_n$ and we look for the width $\Delta z_n$ of the region where the probability distribution $P_n(z)$ exceeds this value, see Fig.~\ref{fig:si-side-peak}. We then use the parameters $\theta_n$ and $\Delta z_n$ to assign a Gaussian with width $\sigma_n$ to the probability distribution $P_n(z)$ according to the relation
\begin{equation}
	\theta_n = \frac{1}{\sqrt{2 \pi \sigma_n^2}} e^{ - \frac{(\Delta z_n/2)^2}{2 \sigma_n^2} },
	\label{eq:thetan}
\end{equation}
obtaining
\begin{equation}
	\sigma_n = \sqrt{- \frac{(\Delta z_n/2)^2}{\textrm{W}_{-1}\left( -2 \pi (\Delta z_n/2)^2 \theta_n^2 \right)} },
\end{equation}
where $W_{-1}$ is the $-1$ branch of the Lambert W function. Considering that the width of the distribution covers multiple orders of magnitude throughout the evolution of the protocol, we need to also choose an adaptive way to determine $\theta_n$, relative to the size of the distribution in each step of the protocol. To do so we rely on the width of the previous prior $\sigma_{n-1}$, which should be close to $\sigma_n$, and set
\begin{equation}
	\theta_n = \frac{1}{ \sqrt{2 \pi \sigma_{n-1}^2}}e^{- \frac{\theta_z^2}{2}},
	\label{eq:si-threshold}
\end{equation}
with $\theta_z$ a parameter that is kept fixed over the duration of the protocol, and which determines the relation between
the threshold value $\theta_n$ and the maximum of the Gaussian distribution. 

Deviations from a Gaussian probability distribution originate in TLS-state down outcomes, whose associated probability distribution displays an additional smaller peak at the tail of the Gaussian, see Fig.~\ref{fig:si-side-peak}. The key advantage of choosing a threshold $\theta_n$ to determine the width (instead of for example computing the variance of $P_n(z)$) is that in this way side peaks can be automatically detected. In the situation depicted in Fig.~\ref{fig:si-side-peak}(a), the side peak is initially below the threshold value, which leads to its growth independently of the measurement outcome, because both inversion profiles $I_n(z|\downarrow)$ and $I_n(z|\uparrow)$ have a significant overlap with it. Eventually the situation depicted in Fig.~\ref{fig:si-side-peak}(b) will arise, where this side peak becomes bigger than $\theta_n$ which consequentially leads to an increased $\Delta z_n$ and $\sigma_{\rm I}$. In this case, the side peak keeps growing for only one of the two possible outcomes. In fact the outcome, where the side peak grows is heavily suppressed because the probability mass covered by the associated inversion profile is small, which is visualized in Fig.~\ref{fig:si-side-peak}(b). However, even in the case of this unlikely outcome the algorithm does not fail, instead the suppression of the side peak is simple deferred to a later measurement. This might lead to a reduction in efficiency of the algorithm, which is why we introduce another trick to suppress the growth of the side peak even when it is still smaller than $\theta_n$. To that end, we flip the sign of the detuning after each TLS-state down measurement, i.e. right after the emergence of the side peak. Therefore, unlike in Fig.~\ref{fig:si-side-peak}(a), the $I_n(z|\uparrow)$ profile does not have an overlap with the side peak and in case of a TLS-state up outcome the side peak gets suppressed.

Our numerical simulations show that with the algorithm described in this section the distribution $P_n(z)$ stays close to a Gaussian whose width is given by $\sigma_n$ and whose mean corresponds to the value of $z$ where $P_n(z)$ is maximal. We show this by computing the Kullback-Leibner divergence between $P_n(z)$ and a Gaussian with the stated parameters. A typical result is shown in Fig.~\ref{fig:si-entropy-kl}(b). As can be seen in this plot, the deviations from the Gaussian form do not accumulate. Furthermore, the Kullback-Leibner divergence is a measure of the distance to a particular target distribution (in this case a specific Gaussian). Therefore, non-zero values of the Kullback-Leibner divergence are not immediately related to deviations from a Gaussian form but can rather mean that the distribution is given by a Gaussian with different parameters than the one we are comparing it to. Hence, it sets a more stringent condition than would be required to prove our point.

The free parameters of our algorithm are the multiplicative factor $w$ in Eq. (\ref{eq:si-update-width}), the target probability $p_\uparrow$ in Eq. (\ref{eq:si-update-detuning}) and the threshold $\theta_z$ in Eq. (\ref{eq:si-threshold}). In the next section, we present the numerical optimization over these parameters.

\section{Optimization of the algorithm parameters}
\begin{figure}[h!]
	\centering
	\includegraphics[width=\textwidth]{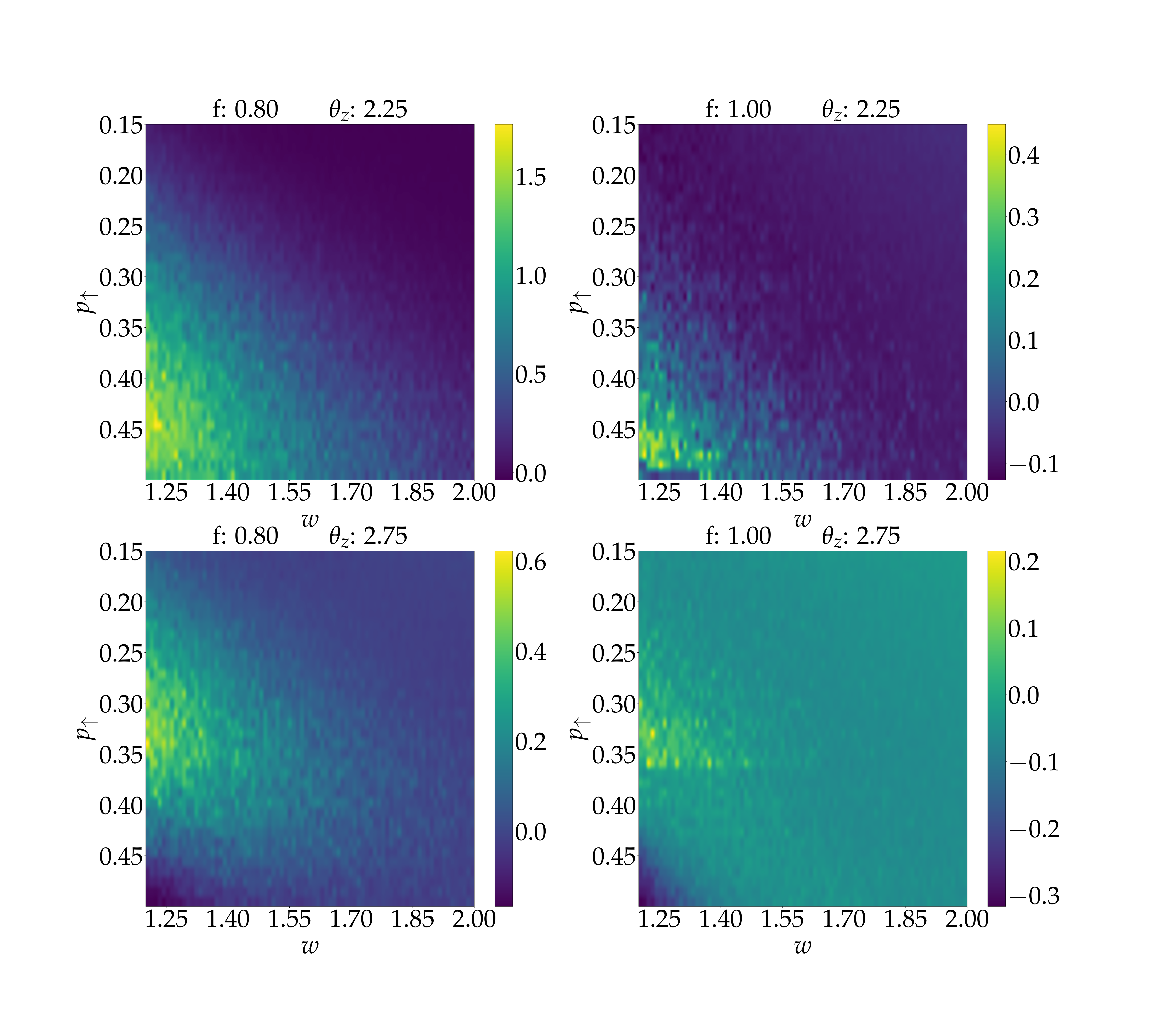}
	\caption{\label{fig:si-final-entropy-optimization} The panels show the final entropy as a function of $w$ and the target probability $p_\uparrow$ for different readout fidelities and threshold values $\theta_z$. While the absolute values of the entropy differ, the scaling with $w$ and $p_\uparrow$ is equal in each panel. The relevant quantity for the performance of the algorithm is the deviation of the entropy from the desired target value rather than the absolute value of the entropy. In the simulations presented here, the target value is the entropy of a Gaussian distribution with width $\sigma=1/2$. This value is subtracted from the value reached by the algorithm, leading to negative values in certain cases. }
\end{figure}

\begin{figure}[h!]
	\centering
	\includegraphics[width=\textwidth]{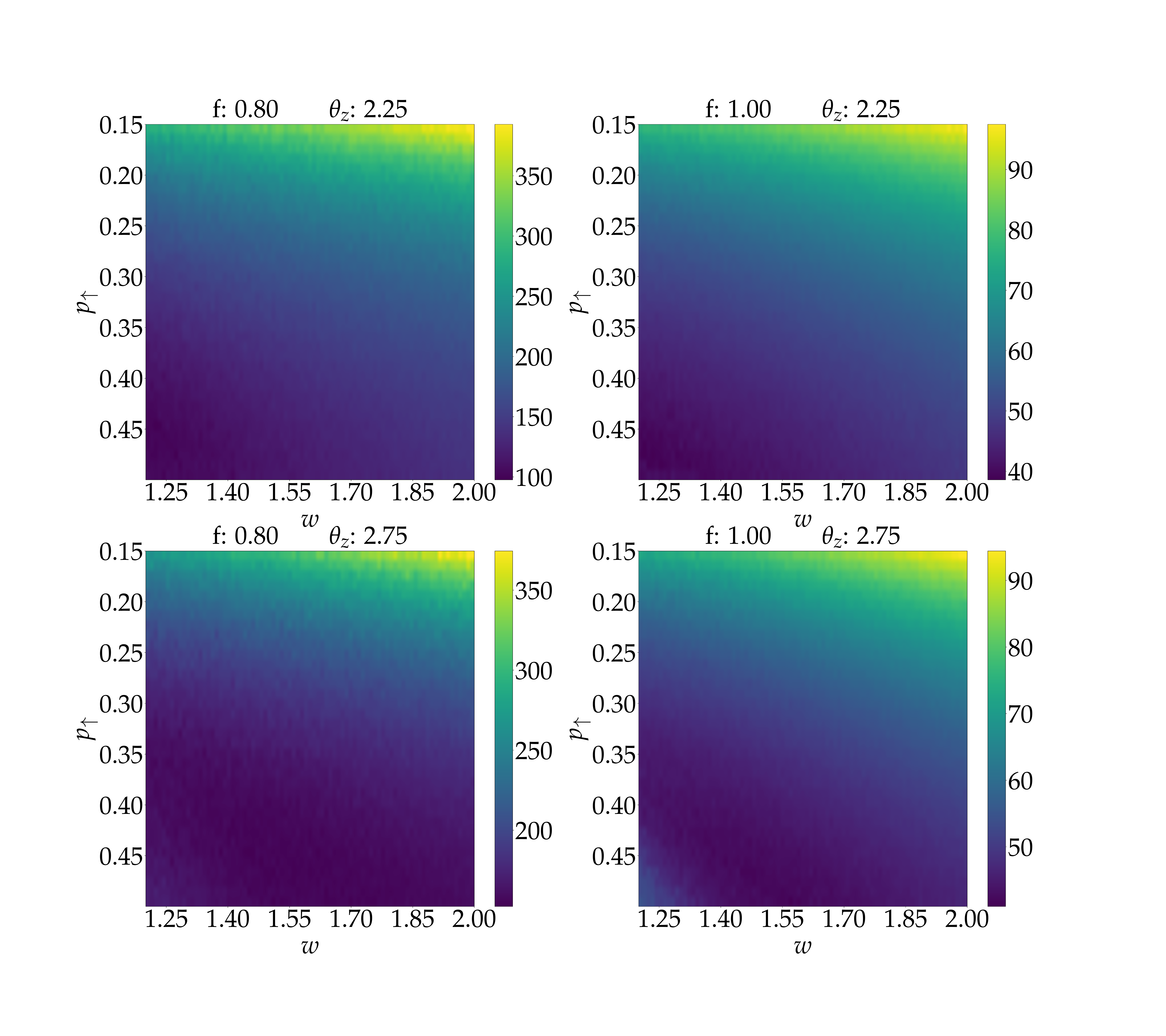}
	\caption{\label{fig:si-protocol-duration-optimization} The panels show the total protocol duration, in units of inverse coupling $1/g$, as a function of $w$ and the target probability $p_\uparrow$ for different readout fidelities and threshold values $\theta_z$. While the absolute values differ, the scaling with $w$ and $p_\uparrow$ is equal in each panel.}
\end{figure}

The adaptive algorithm contains three free parameters: the multiplicative factor $w$ in Eq. (\ref{eq:si-update-width}), the target probability $p_\uparrow$ in Eq. (\ref{eq:si-update-detuning}) and the threshold $\theta_z$ in Eq. (\ref{eq:si-threshold}). The algorithm also requires a stop condition, i.e. the desired state width. We perform simulations of the Bayesian update rule in Eq. (\ref{eq:sl-recursion}) for ranges of these parameters to determine the set that leads to the lowest final entropy in the shortest time. For the stop condition we choose $\sigma = 1/2$. However, this choice does not affect the optimization of the other parameters. We also subtract the entropy of a Gaussian with $\sigma = 1/2$, such that we obtain a final entropy of 0 in the optimal case. However, this subtraction can also lead to negative values of the final entropy. This has no further significance for our analysis as what matters is the deviation from the target value (in this case the entropy of a Gaussian width $\sigma = 1/2$), rather than the absolute value of the final entropy. The results are shown in Figs. (\ref{fig:si-final-entropy-optimization}) and (\ref{fig:si-protocol-duration-optimization}). We see that $w$ and $p_\uparrow$ are independent of the readout fidelity and $\theta_z$. It is also obvious that there is a trade-off between final entropy and protocol duration. The optimal choice of these parameters will depend on the heating rate of the experiment. For the simulations in this work, we chose the parameters $w= 1.9$ and $p_\uparrow=0.4$. We also see that a higher value for $\theta_z$ leads to a significantly lower final entropy while only moderately increasing the protocol duration which is why we chose $\theta_z=2.75$ for our simulations.

\begin{figure}[h]
	\centering
	\includegraphics[width=0.5\textwidth]{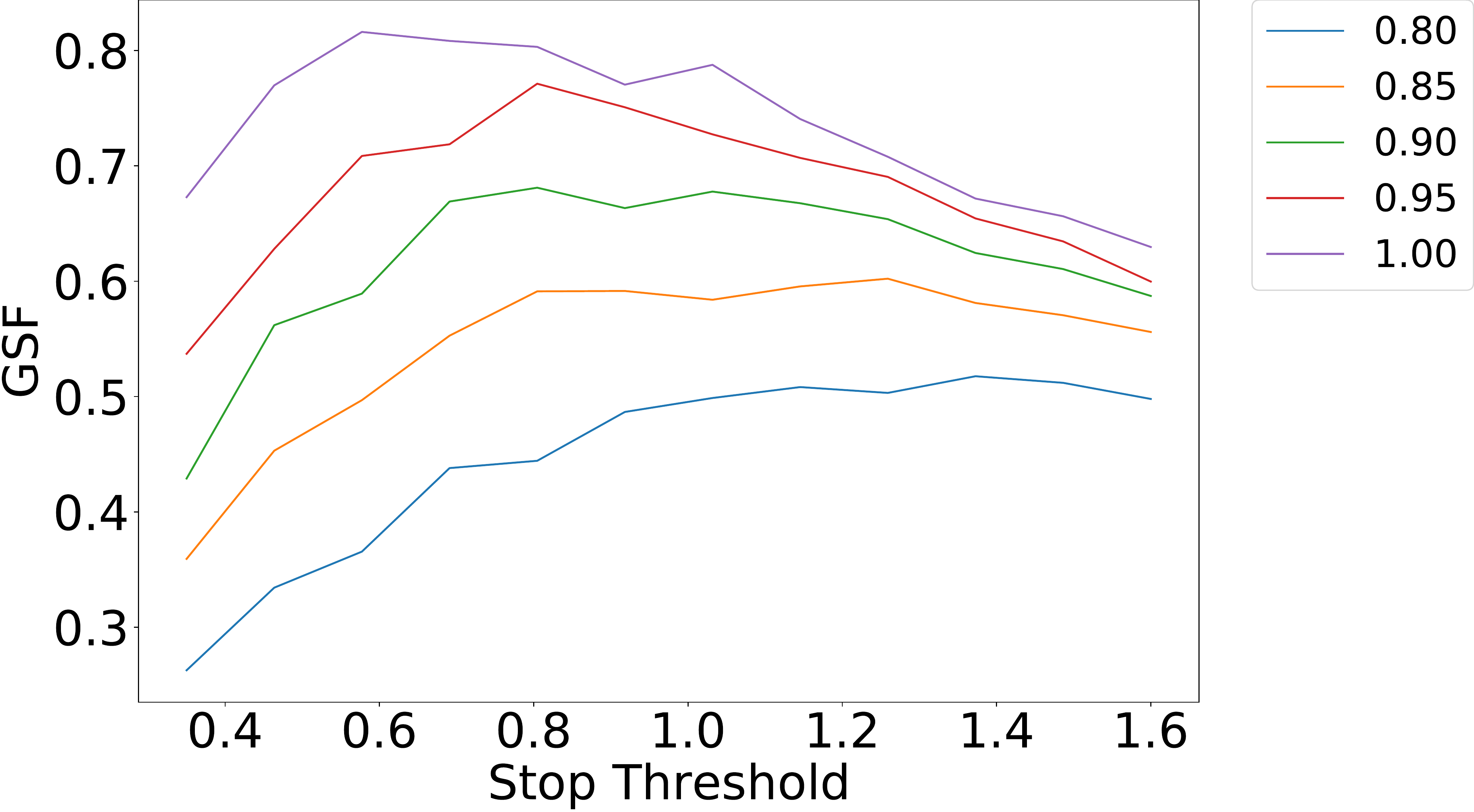}
	\caption{\label{fig:si-threshold-optimization} Ground-state fidelity (GSF) as a function of the stop threshold for the second quadrature plotted for different readout fidelities.}
\end{figure}

After choosing the above parameters, we performed an optimization of the stop condition by simulating the full-quantum dynamics, see Sec.~\ref{sec:simulations} The algorithm requires two stop conditions, one for each quadrature. We observed in our numerical simulations that the GS fidelity increases with the amount of squeezing that is introduced in the first quadrature. Therefore, a threshold on $\sigma_n$, that is related to the width of the probability distribution via Eq.~(\ref{eq:thetan}), should be chosen that achieves the maximal amount of squeezing, while considering the heating rate of the particle and the dephasing rate of the TLS. These set a bound on the amount of squeezing that can be achieved. We numerically optimized the threshold for the second quadrature and the results are shown in Fig.~\ref{fig:si-threshold-optimization}. However, the heating rates were set to zero in these simulations. The optimal threshold depends on the readout fidelity $f$ and comes close to the expected value of the coherent state width $\sigma = 1/\sqrt{2}$ for $f=1$.

\newpage
\section{Simulations}
\label{sec:simulations}
In this section we present the full-quantum simulation that we used to generate the data presented in Fig.~3 of the main text. To validate our algorithm we numerically integrated the Lindblad equation
\begin{align*}
	\dot{\hat{\rho}} &= -i \left[\hat{H}(t), \hat{\rho} \right] + \mathcal{L}(\hat{\rho}), \\
	\hat{H}(t) &= \frac{g}{2} \hat{z} \hat{\sigma}_z+ \frac{\Omega(t)}{2}\hat{\sigma}_x, \\
	\mathcal{L}(\hat{\rho}) &= \frac{\gamma}{2} \left(\bar{n} + \frac{1}{2} \right) \left( 2 \hat{z}\hat{\rho}\hat{z} - \hat{z}^2 \hat{\rho} - \hat{\rho} \hat{z}^2 + 2 \hat{p} \hat{\rho} \hat{p} - \hat{p}^2 \hat{\rho} - \hat{\rho} \hat{p}^2 \right) \\
	& + i \frac{\gamma}{4} ( 2 \hat{p} \hat{\rho} \hat{z} - 2 \hat{z} \hat{\rho} \hat{p} + \hat{p}\hat{z}\hat{\rho} + \hat{\rho}\hat{p}\hat{z} - \hat{z}\hat{p}\hat{\rho} - \hat{\rho}\hat{z}\hat{p}).
\end{align*}
Here, we set $\hbar = \sqrt{\hbar/(m \omega_M)} = 1$. To be able to efficiently simulate high-temperature states, we reformulated these equations in the phase space of the levitated particle while keeping the Hilbert space formulation for the TLS. This was done by writing the density matrix as
\begin{equation}
	\hat{\rho} = W_{\uparrow \uparrow} \ketbra{\uparrow} + W_{\uparrow \downarrow} \ketbra{\uparrow} + W_{\downarrow \uparrow} \ketbra{\downarrow} + W_{\downarrow \downarrow} \ketbra{\downarrow}.
\end{equation}
Here, $W_{ij}$ are Wigner functions multiplied with the probability amplitude of the associated TLS state. We define the Wigner functions via the characteristic function \cite{carmichael-statistical-methods-1}
\begin{align*}
	\chi(\xi, \eta) = \textrm{Tr}\left( \rho e^{i \xi \hat{z} + i \eta \hat{p}}  \right) = \int_{-\infty}^{\infty} \textrm{d}\mu \int_{-\infty}^{\infty} \textrm{d}\nu W(\mu, \nu) e^{i \xi \mu} e^{i \eta \nu}.
\end{align*}
This leads to the following equations of motion
\begin{align}
	\dot{W}_{\uparrow \uparrow}(\mu,\nu) &= -i \left( \frac{\Omega}{2} W_{\downarrow \uparrow}(\mu,\nu) - \frac{\Omega}{2} W_{\uparrow \downarrow}(\mu,\nu) \right) + \frac{g}{2} \frac{\partial}{\partial \nu}W_{\uparrow \uparrow}(\mu,\nu) + \mathcal{D}\left[W_{\uparrow \uparrow}\right], \\
	\dot{W}_{\uparrow \downarrow}(\mu,\nu) &= -i \left( \frac{\Omega}{2} W_{\downarrow \downarrow}(\mu,\nu) - \frac{\Omega}{2} W_{\uparrow \uparrow}(\mu,\nu) \right) -i g \mu W_{\uparrow \downarrow}(\mu,\nu) + \mathcal{D}\left[W_{\uparrow \downarrow}\right], \\
	\dot{W}_{\downarrow \uparrow}(\mu,\nu) &= -i \left( \frac{\Omega}{2} W_{\uparrow \uparrow}(\mu,\nu) - \frac{\Omega}{2} W_{\downarrow \downarrow}(\mu,\nu)\right) + ig \mu W_{\downarrow \uparrow}(\mu,\nu) + \mathcal{D}\left[W_{\downarrow \uparrow}\right], \\
	\dot{W}_{\downarrow \downarrow}(\mu,\nu) &= -i \left( \frac{\Omega}{2} W_{\uparrow \downarrow}(\mu,\nu) - \frac{\Omega}{2} W_{\downarrow \uparrow}(\mu,\nu)\right) - \frac{g}{2} \frac{\partial}{\partial \nu} W_{\downarrow \downarrow}(\mu,\nu) + \mathcal{D}\left[W_{\downarrow \downarrow}\right], \\
	\mathcal{D}\left[ W \right] &= \left[ \frac{\gamma}{2} \left( \frac{\partial}{\partial \mu} \mu + \frac{\partial}{\partial \nu}\nu \right) + \frac{\gamma}{2}\left(\bar{n} + \frac{1}{2} \right) \left(\frac{\partial^2}{\partial \mu^2} + \frac{\partial^2}{\partial \nu^2}\right) \right] W.
\end{align}
These equations contain the heating rate $\Gamma$ from the main text by the relationship $\Gamma = \gamma \bar{n}$. The These simulations serve to validate the algorithm, but are not required to determine the measurement parameters in each step. 

The algorithm requires a prior distribution, which for thermal states and with our conventions corresponds to the Gaussian
\begin{equation}
	P_0(z) = \frac{1}{\sqrt{2 \pi \sigma^2}} e^{ - \frac{z^2}{2\sigma^2}},
\end{equation}
with $\sigma = \sqrt{ \bar{n} + 1/2}$. We quantify the distance to the GS via the fidelity
\begin{equation}
	\mathcal{F}(\rho, \sigma) = \textrm{Tr}\left[\sqrt{ \sqrt{\sigma} \rho \sqrt{\sigma}}\right]^2 = \bra{0}\rho\ket{0} = 2\int \textrm{d}\mu \int \textrm{d}\nu W(\mu, \nu) e^{-\mu^2} e^{-\nu^2}.
\end{equation}

\section{Derivation of the Bayesian update rule with imperfect readout}
In this section we present the derivation of the Bayesian update rule in Eq. (3) of the main text, which accounts for imperfect readout fidelities. We define the readout fidelity as the probability of correctly detecting a given TLS state, i.e. we do not account for errors in state preparation. Furthermore we assume that the readout fidelity is equal for both states. This can be expressed by the conditional probabilities:
\begin{align}
	p\left(1 \given \uparrow \right) &= f, \quad p\left(0 \given \uparrow \right) = 1-f, \\
	p\left(0 \given \downarrow \right) &= f, \quad p\left(1 \given \downarrow \right) = 1-f.
\end{align}
Where $p\left(1 \given \uparrow \right)$ ($p\left(0 \given \downarrow \right)$) is the conditional probability that the TLS-up (down) state is detected if the TLS is in the up (down) state. The spatial probability distribution after up state detection is given by
\begin{equation}
	P\left(z \given 1 \right) = \sum_{s=\{\uparrow, \downarrow\}} P \left(z,s \given 1 \right) = \sum_{s=\{\uparrow, \downarrow\}} P\left(z \given s,1 \right) p\left(s \given 1\right) = \sum_{s=\{\uparrow, \downarrow\}} \frac{p\left(1 \given s\right) p\left(s\right)}{p\left(1\right)} P\left(z \given s,1 \right).
	\label{eq:sp-1}
\end{equation}
Furthermore, we know that $P \left(z \given s, 1 \right) \equiv P \left(z \given s\right)$ and that the conditional spatial distribution for spin state $s=\{\uparrow, \downarrow\}$ is given by
\begin{equation}
	P \left(z \given s, 1 \right) = \frac{1}{p\left(s\right)} \left| \bra{s} U'(z) \ket{\downarrow} \right|^2 P\left( z \right),
	\label{eq:sp-s1}
\end{equation}
as derived in Sec. \ref{sec:ShapedInversionProfiles}, and
\begin{equation}
	p\left( s \right) = \int_{-\infty}^\infty \mathrm{d}z \left| \bra{s} U'(z) \ket{\downarrow} \right|^2 P\left( z \right).
\end{equation}
Here and in the following $P(z)$ is the spatial probability distribution of the initial state of the oscillator. Plugging Eq. (\ref{eq:sp-s1}) into Eq. (\ref{eq:sp-1}) yields the spatial probability distribution given the measurement result 1
\begin{equation}
	P\left( z \given 1 \right) = \frac{1}{p\left( 1 \right)} \left( f \left| \bra{\uparrow} U'(z) \ket{\downarrow}\right|^2 + (1-f) \left| \bra{\downarrow} U'(z) \ket{\downarrow}\right|^2 \right) P\left(z\right),
\end{equation}
with
\begin{equation}
	p\left( 1 \right) = p\left( 1 \given \uparrow \right) p\left(\uparrow\right) + p\left(1 \given \downarrow \right) p\left( \downarrow \right) = f p\left(\uparrow \right) + (1-f) p\left( \downarrow \right)
\end{equation}

\end{document}